\documentclass{article}  

\usepackage{graphicx}

\usepackage{geometry}
\usepackage{graphicx}
\usepackage{natbib}
\usepackage{color}
\usepackage{amssymb}
\usepackage{lineno}
\usepackage{txfonts}
\usepackage{amstext}
\usepackage{mathabx}
\usepackage{multirow}
\usepackage{ulem}
\usepackage{lscape}
\usepackage{txfonts}
\usepackage{subfigure}
\usepackage{float}

\usepackage{hyperref}

\newcommand{\Kunit}{\,cm$^{2}\cdot$s$^{-1}$}

\newcommand{\fig}[1]{Fig.~\ref{#1}}

\begin{document} 


\Large
\begin{center}\textbf{The deep oxygen abundance in Solar System Giant Planets, with a new derivation for Saturn}\end{center}\normalsize

\large\noindent T. Cavali\'e$^{1,2}$, J. Lunine$^{3}$, O. Mousis$^{4}$, R. Hueso$^{5}$\normalsize\\
\vspace{0.2cm}

\noindent$^1$Laboratoire d'Astrophysique de Bordeaux, Univ. Bordeaux, CNRS, B18N, all\'ee Geoffroy Saint-Hilaire, 33615 Pessac, France (ORCID: 0000-0002-0649-1192, email: thibault.Cavali\'e@u-bordeaux.fr)\\ 
$^2$LESIA, Observatoire de Paris, PSL Research University, CNRS, Sorbonne Universit\'es, UPMC Univ. Paris 06, Univ. Paris Diderot, Sorbonne Paris Cit\'e, Meudon, France\\
$^3$ Cornell University, Department of Astronomy, Ithaca NY, USA\\
$^4$ Aix-Marseille Universit\'e, CNRS, CNES, Institut Origines, LAM, Marseille, France\\
        Institut Universitaire de France (IUF), France\\
$^5$ Departamento de F\'isica Aplicada, Escuela de Ingenier\'ia de Bilbao, Universidad del Pa\'is Vasco/Euskal Herriko Unibertsitatea UPV/EHU, Plaza Ingeniero Torres Quevedo, 1, E-48013 Bilbao, Spain\\

\vspace{0.2cm}
\noindent\textbf{Received:} 26 September 2023\\
\noindent\textbf{Accepted:} 8 January 2024\\
\noindent\textbf{Published:} 22 January 2024\\
\vspace{0.2cm}

\noindent\textbf{DOI:} https://doi.org/10.1007/s11214-024-01045-6\\
\vspace{0.5cm}

\section*{Abstract}
Deep elemental composition is a challenging measurement to achieve in the giant planets of the solar system. Yet, knowledge of the deep composition offers important insights in the internal structure of these planets, their evolutionary history and their formation scenarios. A key element whose deep abundance is difficult to obtain is oxygen, because of its  propensity for being in condensed phases such as rocks and ices. In the atmospheres of the giant planets, oxygen is largely stored in water molecules that condense below the observable levels. At atmospheric levels that can be investigated with remote sensing, water abundance can modify the observed meteorology, and meteorological phenomena can distribute water through the atmosphere in complex ways that are not well understood and that encompass deeper portions of the atmosphere. The deep oxygen abundance provides constraints on the connection between atmosphere and interior and on the processes by which other elements were trapped, making its determination an important element to understand giant planets. In this paper, we review the current constraints on the deep oxygen abundance of the giant planets, as derived from observations and thermochemical models.

\section{Introduction}
\label{intro}
Giant planet formation and evolution remain outstanding questions in planetary science to address more generally the formation of planetary systems. Measurements and models of giant planet gravity (Folkner et al. 2017, Iess et al. 2019, Movshovitz et al. 2022), luminosity (Hanel et al. 1981, 1983, Pearl et al. 1990, Pearl \& Conrath 1991) and magnetic fields (Connerney et al. 1987, 1991, 2018, 2022, Dougherty et al. 2018, Ness et al. 1989) combined with deep composition and isotopic ratios (Niemann et al. 1998, Mahaffy et al. 2000, Wong et al. 2004, Lellouch et al. 2001, Feuchtgruber et al. 2013) to derive their internal structure hold part of the key to understand the formation of these planets (Helled et al. 2011, 2020, Helled \& Fortney 2020, Nettelmann et al. 2013, Guillot et al. 2022). The amount of heavy elements incorporated in their cores and envelopes structures their interior and governs their long-term evolution history (Guillot 2005, Nettelmann et al. 2016, Friedson et al. 2017).

Noble gases are key diagnostics of formation models (e.g., Atreya et al. 1999, Mousis et al. 2018) but can only be measured with in situ probes (Mahaffy et al. 2000), with the exception of helium which can be derived from continuum measurements sensitive to the H$_2$-He collision-induced absorption spectra (Gautier et al. 1981, Conrath et al. 1987, 1991, 2000, Koskinen \& Guerlet 2018). However, the level of accuracy necessary to constrain the models requires in situ measurements, that are achievable with a mass spectrometer and/or an interferometer experiment (Niemann et al. 1998, von Zahn \& Hunten 1992, vonZahn et al. 1998). 

Other heavy elements can, in principle, be derived from remote sensing observations of their main carrier molecule, which are CH$_4$, NH$_3$, H$_2$S, PH$_3$, H$_2$O, etc. in the reducing atmospheres of the giant planets. Such observations are carried out at various wavelengths ranging from the near infrared to the radio. The abundance measurements are achieved from reflection spectroscopy or thermal emission (e.g., de Pater et al. 1989, 1991, 2023, Fletcher et al. 2007, Fletcher et al. 2009, Irwin et al. 2018, 2019, Li et al. 2020, Molter et al. 2021, Li et al. 2023a). For example, given its high volatility, gaseous methane observations in all four giant planets have constrained the deep carbon abundance, albeit with lower accuracy at Uranus and Neptune (Karkoschka \& Tomasko 2009, 2011, Sromovsky \& Fry 2008, Sromovsky et al. 2011, 2014, Irwin et al. 2019, 2021) than at Jupiter and Saturn (Wong et al. 2004, Fletcher et al. 2009). 

The abundance of oxygen and the C/O ratio in the envelope are of particular interest as they tell us about the amount of ices incorporated in the planet envelope. They can then be tied to where and when the planet formed in the protosolar nebula (PSN) with the caveat that mixing between the observable atmosphere and interior may be incomplete (Mousis et al. 2012, 2019, 2020, 2021, Schneider \& Bitsch 2021, Aguichine et al. 2022, Schneeberger et al. 2023, Pacetti et al. 2022). In addition, the deep oxygen abundance is also diagnostic of ice condensation processes and of how other volatiles were trapped to later enrich the planet envelope (Lunine \& Stevenson 1985, Bar-Nun et al. 1988, Owen et al. 1999, Gautier et al. 2001, Hersant et al. 2004,Helled \& Lunine 2014). Finally, it also bears implications on evolution through its influence on the thermal profile, by producing radiative layers at the various condensation levels (Guillot 1995) such as that of water when the deep oxygen mixing ratio is higher than 7\% (Leconte et al. 2017, Cavali\'e et al. 2017), a condition that is particularly relevant to Uranus and Neptune.

There are, however, several difficulties in the derivation of accurate enough abundances from remote sensing observations. First, there is the degeneracy with tropospheric temperatures, which often needs to be extrapolated from higher levels where temperatures are measured, to explain the observed spectral features. Second, some of these species can be locked by condensation at levels that are too deep for remote sensing to be sensitive enough. The more distant and colder the planet, the more affected the species. This is particularly the case for H$_2$O in the ice giants (Atreya et al. 2020). Finally, even probing below their respective cloud levels does not ensure probing the well-mixed layers from which the deep abundances of thee species can be derived, as exemplified by the observations of NH$_3$ and H$_2$O in Jupiter with the Galileo Probe in a single location (Atreya et al. 2020) with specific dynamics (Showman \& Dowling 2000), or from global maps of NH$_3$ in Jupiter obtained with Juno (Bolton et al. 2017, Li et al. 2017). Thermochemical modeling work can then complement these remote sensing observations, especially by using species not affected by condensation in the atmospheres of the giant planets as probes to the deep elemental composition (e.g., Prinn \& Barshay 1977, Fegley et al. 1988, Yung et al. 1988, Lodders \& Fegley1994, Visscher et al. 2005, 2010, Cavali\'e et al. 2017, Venot et al. 2020). This is particularly the case of the CO molecule, a disequilibrium species observed in all giant planets (Beer 1975, B\'ezard et al. 2002, Bjoraker et al. 2018, Noll et al. 1986, Fouchet et al. 2017, Encrenaz et al. 2004, Cavali\'e et al. 2014, Marten et al. 1993, Lellouch et al. 2005) and used to probe their deep oxygen abundance.

In this paper, we will review the current state of knowledge regarding the deep oxygen abundance in giant planets, i.e. well below the 100-bar level where condensates and meteorology affect the abundances, as constrained from in situ measurements, remote sensing observations and models. We will discuss the implications of oxygen-rich and oxygen-depleted envelopes on the formation and evolution of the giant planets of the solar system.

\section{Oxygen-rich or oxygen-poor envelopes?}
\label{Envelopes}

  \subsection{Carbon abundance in giant planets} \label{Carbon}
  Whether a giant planet is oxygen-rich or oxygen-poor can be assessed by using the O/H or C/O ratios in their deep atmospheres as a diagnostic. Consequently, and before presenting our current knowledge on the deep oxygen abundance of the giant planets, we thus shortly summarize our knowledge of their deep carbon abundance.
  
  The deep carbon abundance can be retrieved from in situ or remote sensing observations of methane, which is the main carbon carrier in the reducing atmospheres of the giant planets. In what follows, we use the following measurements: Galileo probe for Jupiter (Wong et al. 2004), Cassini-CIRS for Saturn (Fletcher et al. 2009), HST observations for Uranus (Sromovsky et al. 2011), and a combination of HST and VLT observations for Neptune (Karkoschka \& Tomasko 2011, Irwin et al. 2019). Thermochemical modeling can help to constrain the C abundance deeper than the levels where it is measured. This is the case in Cavali\'e et al. (2023) for Jupiter, in this paper for Saturn (see Section \ref{Saturn_new}), and in Venot et al. (2020) for Uranus and Neptune. We note that, because the methane abundance is not uniform with latitude below the methane cloud in the ice giants (e.g., Sromovsky et al. 2011, Irwin et al. 2021), the equatorial methane abundance is assumed to be representative of the deeper well-mixed layers for the ice giants in those models. This surely prompts for the need of a better knowledge of the spatial variations of methane both as a function of latitude and altitude in ice giant atmospheres. This could be achieved with a a set of 2+ probes entering their atmospheres, as proposed for Uranus (Wong et al. 2024), with the aim of reaching a minimum pressure of 5 bars.
  
  In the next sections, we use the following C/H ratios with respect to the protosolar value of Lodders 2021): 3.28$\pm$0.81 for Jupiter, 7.30$\pm$0.43 for Saturn, 59$\pm$20 for Uranus, and 45$\pm$20 for Neptune.

  \subsection{Constraints from oxygen species observations and thermochemical models}
  \subsubsection{Jupiter}
  Jupiter has been the most intensively studied planet with respect to the determination of its deep oxygen abundance. The initial detection of CO by Beer (1975) at abundance levels orders of magnitude higher than predicted by thermochemical equilibrium models raised the possibility that Jupiter had a significant deep water reservoir and sufficiently rapid convection to quench the conversion of CO into H$_2$O and CH$_4$ at depth and maintain the CO abundance high enough to make it observable (Prinn \& Barshay 1977).

  The use of quench level models, in which the full thermochemistry is approximated by a pre-identified rate-limiting reaction, enabled Fegley et al. (1988) to further constrain the deep oxygen abundance of Jupiter from CO observations. In the past two decades, CO has been re-observed in the 5-$\mu$m window and B\'ezard et al. (2002) found that CO had a tropospheric (4-8 bar) mole fraction of 1.0$\pm$0.2~ppb. The higher altitude increase of the CO abundance is caused by external sources of oxygen (B\'ezard et al. 2002, Moses \& Poppe 2017), among which the Shoemaker-Levy 9 has been identified as the main contributor (Lellouch et al. 1995, 2002, Cavali\'e et al. 2013). Bjoraker et al. (2018) confirmed the results of B\'ezard et al. (2002) by deriving an upper tropospheric CO mole fraction of 0.8$\pm$0.2~ppb. With the rate-limiting reaction proposed by Yung et al. (1988), B\'ezard et al. (2002) found that the CO observations could be matched with a deep oxygen abundance of 0.2-9 times solar. This result is not accurate enough to discriminate between the two models for ice condensation and heavy element trapping processes, i.e. adsorption on amorphous ice (Bar-Nun et al. 1988, Owen et al. 1999) and clathrates (Lunine \& Stevenson 1985, Gautier  et al. 2001, Gautier \& Hersant 2005). 
  
  Comprehensive thermochemical and diffusion models have been developed ever since. Visscher et al. (2010) derived a deep oxygen abundance of 0.3-7.3 times the solar value. Wang et al. (2015) compared the chemical schemes of Visscher \& Moses (2011) and Venot et al. (2012) and found oxygen abundances of 0.1-0.75 and 3-11 times the solar value, respectively. This rather significant discrepancy between the results, also confirmed by Wang et al. (2016), had been identified as originating from the methanol chemistry by Moses (2014). Venot et al. (2020) undertook a full revision of their chemical scheme and the most recent determination by Cavali\'e et al. (2023) of the deep oxygen abundance with their model confirms the sub-solar oxygen abundance in Jupiter (0.3$^{+0.5}_{-0.2}$ times the protosolar value) and a possible C/O ratio of 6$^{+10}_{-5}$, i.e. 12$^{+20}_{-10}$ times the protosolar value.
  
  More direct measurements of the tropospheric water abundance, as a direct probe of the deep oxygen, have been attempted. After the first detection of H$_2$O in Jupiter's troposphere by Larson et al. (1975), Kunde et al. (1982) was able to constrain the H$_2$O abundance for the first time in Jupiter's troposphere. However, the fact that they found increasing mixing ratio with depth showed that their measurements probed the saturation curve and did not reach the well-mixed region. This was again the case for Encrenaz et al. (1996) with the Infrared Space Observatory, Nixon et al. (2001) with Galileo/NIMS and Bjoraker et al. (2022) with Keck. The Galileo probe, which descended into Jupiter atmosphere on December 7, 1995, measured an H$_2$O abundance down to the 22-bar level with its mass spectrometer and found it substantially depleted with respect to the Sun (Niemann et al. 1998, Wong et al. 2004). This was confirmed by the Net Flux Radiometer which constrained the water vapor to $<$6\% solar at 10 bars (Sromovsky et al. 1998). However, later Juno data indicated that the probe may not have reached the well-mixed region. The H$_2$O abundance measured by the mass spectrometer was indeed still increasing at loss-of-signal at $\sim$22 bars and the NFR measurements were consistent with a water abundance increasing with depth. This was later attributed to local meteorology (Showman \& Dowling 2000), with the probe entering a 5-$\mu$m hot spot (Orton et al. 1998). The oxygen abundance of 0.289$\pm$0.096 times the solar value (Wong et al. 2004) measured by Galileo may thus be a lower limit. 

  Finally, the Juno spacecraft recorded Jupiter's microwave emissions over multiple emission angles to remove the degeneracy between NH$_3$ and H$_2$O absorption. Li et al. (2020) isolated the data pertaining to the equatorial region, where NH$_3$ was found well-mixed up to its cloud level (Bolton et al. 2017, Li et al. 2017), and derived a nominal H$_2$O abundance of 1.0-5.1 times the protosolar value of Asplund et al. (2009). It must however be noted that, if Juno favors solar to supersolar H$_2$O, the 2-$\sigma$ error bar encompasses subsolar to supersolar abundances (0.1-7.5 times the protosolar value of Asplund et al. 2009). It should be noted that, when allowing both temperature and composition to be retrieved simultaneously, Li et al. (2022, 2023b) now find a superadiabatic temperature profile in the equatorial region, which requires stabilization from a supersolar water molecular weight gradient.

  \subsubsection{Saturn} \label{Saturn}
  Direct observation of tropospheric water in Saturn is more difficult than on Jupiter, because of Saturn's colder tropopause. Like on Jupiter, ISO observations did not enable to reach the well-mixed layers and de Graauw et al. (1997) found H$_2$O to be undersaturated under 3 bars. Saturn's centimeter-wave spectrum is nonetheless sensitive to opacity caused by the deep water. Interferometric observations (de Pater \& Dickel 1991, Dunn et al. 2005) can be reproduced with roughly 5 times solar H$_2$O and more recent metric observations carried out with the Giant Metrewave Radio Telescope (GMRT) hinted at a deep oxygen abundance of at least 15 times solar (Courtin et al. 2015). However, these estimates remain uncertain because of limitations in absolute calibration of these measurements and of degeneracy with other opacity sources (mainly NH$_3$).
  
  The initial detection of CO in Saturn's atmosphere by Noll et al. (1986) was an indication of an oxygen-rich envelope, but the authors underlined that their results were also fully consistent with a depleted envelope if the CO source was from the exterior of the planet. Follow-up observations of Noll et al. (1991) still lacked the spectral resolution to differentiate an internal source from an external one. Combining these CO observations from the 1990s and abundance measurements of SiH$_4$ and PH$_3$ with thermochemical kinetic calculations, Visscher et al. (2005) established that Saturn's deep oxygen abundance should be 3.2-6.4 times the solar value. Because carbon is 7.30$\pm$0.43 times protosolar (see Section\ref{Carbon}), this oxygen abundance range is indicative of a supersolar C/O ratio in the planet. In the light of more recent observations of CO, we can now update these numbers.

  \subsubsection{A new derivation for Saturn} \label{Saturn_new}
\begin{figure}[!h]
    \begin{center}
      \includegraphics[width=\textwidth,keepaspectratio]{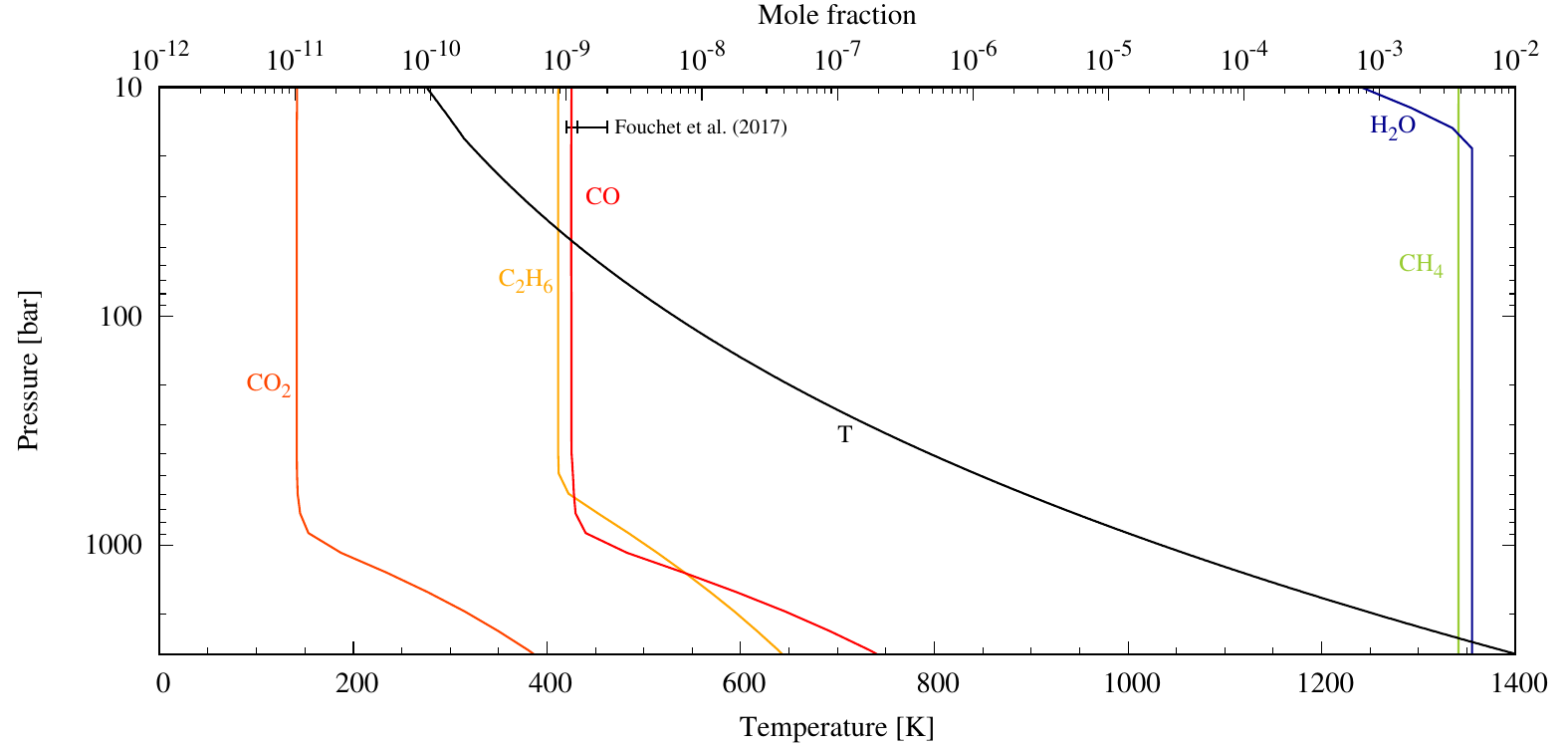}
    \end{center}
    \caption{New derivation of Saturn's upper tropospheric composition as obtained from thermochemical and diffusion simulations using the model of Cavali\'e et al. (2023). The temperature profile used in the calculations is displayed in black. The main species involved in the thermochemical equilibrium between carbon and oxygen are shown. The deep oxygen abundance required to reproduce the CO observations of Fouchet et al. (2017) is nominally 8 times the protosolar value of Lodders (2021).}
    \label{Saturn_oxygen} 
  \end{figure}

  Millimeter observations by Cavali\'e et al. (2009, 2010) enabled to prove the existence of an external source of CO and Fouchet et al. (2017) constrained the internal CO to be nominally 1.2~ppb, with a 1-$\sigma$ range of 1-2~ppb from VLT/CRIRES observations. We can use this recent measurement of the CO tropospheric abundance to further constrain Saturn's deep oxygen abundance. Adapting the model of Cavali\'e et al. (2023) to Saturn's conditions, using a constant vertical mixing coefficient of 10$^8$\Kunit, we can reproduce the upper tropospheric CO of Fouchet et al. (2017). We find that the deep oxygen is nominally 8 times the protosolar value of Lodders (2021), with a range of 7-15 times the protosolar value when accounting for the full 1$\sigma$ range of Fouchet et al. (2017). These outcomes are in agreement with results from Wang et al. (2016) and the nominal simulation results are shown in \fig{Saturn_oxygen}. Saturn's envelope therefore seems to be enriched in oxygen with respect to carbon. We note that we have assumed there was no deep radiative layer in Saturn's troposphere at the location of the CO quench level (such a case is presented and discussed for Jupiter in Cavali\'e et al. 2023). If such layer would exist, then the deep oxygen could be even larger.

  \subsubsection{Uranus and Neptune}
  The first detection of CO in the ice giants was achieved by Marten et al. (1993) and confirmed by Rosenqvist et al. (1992) in Neptune. Surprisingly, they found 1~ppm of CO in the planet, i.e. 3 orders of magnitude more than in Jupiter and Saturn. Combined with the high luminosity of the planet (Pearl \& Conrath 1991), this was taken as solid evidence for an active deep convection-driven meteorology. Lodders \& Fegley (1994) constrained the deep oxygen to 440 times the solar value. However, Lellouch et al. (2005) later identified that CO had a strong external component, which was confirmed by observations of Hesman et al. (2007), Luszcz-Cook \& de Pater (2013) and Teanby et al. (2019). Refining the magnitude of the external source resulted in a decrease of the magnitude of the internal source of CO, up to the limit that some models do not even require an internal source to match the data (Luszcz-Cook \& de Pater 2013, Teanby et al. 2019). The current nominal value of the tropospheric CO abundance is 0.2~ppm and Venot et al. (2020) derived a deep oxygen abundance of 200 times the protosolar value of Lodders (2021). With carbon at 45 times the protosolar value, Neptune's C/O ratio is nominally subsolar. However, any lower oxygen abundance, and thus higher C/O ratio, still seems possible according to the data, making Neptune a planet potentially more rocky than icy (Teanby et al. 2020).

  In Uranus, the situation is still fuzzy. Although CO was detected nearly two decades ago by Encrenaz et al. (2004), the internal source was never conclusively detected (Teanby \& Irwin 2013). Only an external source is firmly established (Cavali\'e et al. 2014). This may result from deep inhibition of convection, as hinted by the lack of luminosity of the planet (Pearl et al. 1990). Taking the upper limit set by Teanby \& Irwin (2013), Venot et al. (2020) derived from this value an oxygen upper limit of 37 times the protosolar value of Lodders (2021). With carbon 60 times protosolar in their model results, the C/O ratio of Uranus is nominally supersolar, but its true deep value depends on the presence of a convection inhibition boundary deeper in the planet.

  \subsubsection{Summary from oxygen species observations and thermochemical models}
\begin{figure}[!h]
    \begin{center}
      \includegraphics[width=10cm,keepaspectratio]{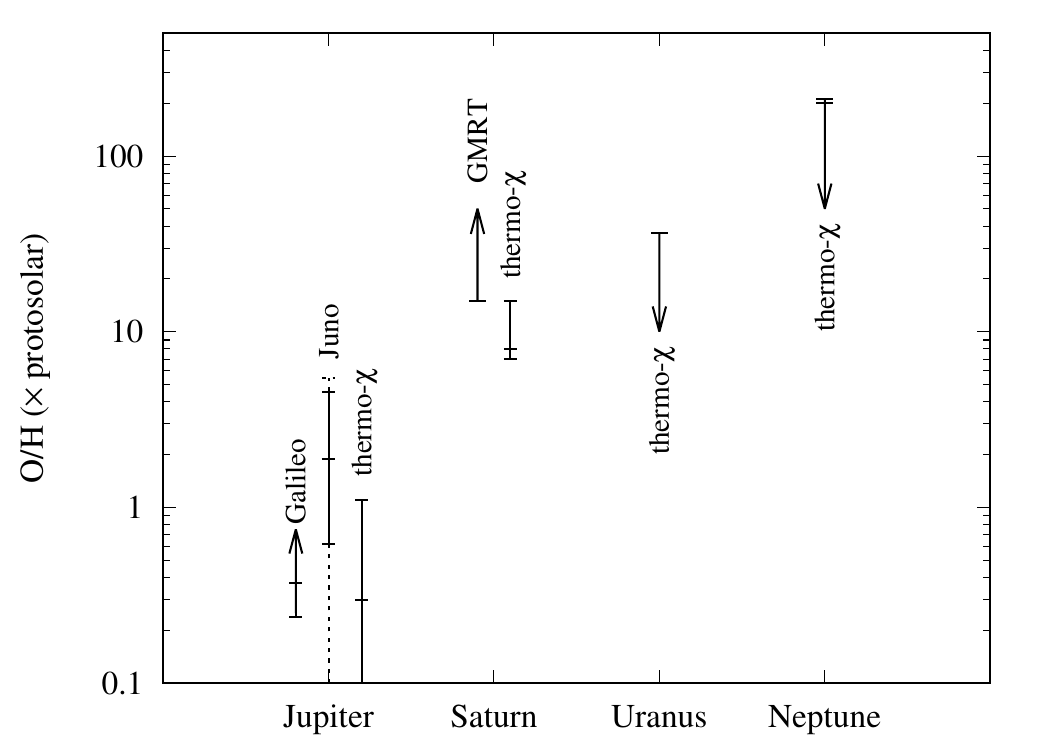}
    \end{center}
    \caption{Oxygen abundances as derived in the four giant planets from various observations of oxygen species and thermochemical models. The oxygen abundances are expressed with respect to the protosolar oxygen abundance of Lodders (2021). For Jupiter, Galileo results are from Wong et al. (2004) and may not be representative of the deep interior (Atreya et al. 1999), Juno results (with 1 and 2-$\sigma$ error bars) from Li et al. (2020), thermochemical results from Cavali\'e et al. (2023). For Saturn, results from GMRT observations are from Courtin et al. (2015) and thermochemical model results are from this paper. For Uranus and Neptune, thermochemical model results are from Venot et al. (2020), accounting for the results of Teanby et al. (2020).}
    \label{fig:abundances} 
  \end{figure}

\begin{figure}[!h]
    \begin{center}
      \includegraphics[width=10cm,keepaspectratio]{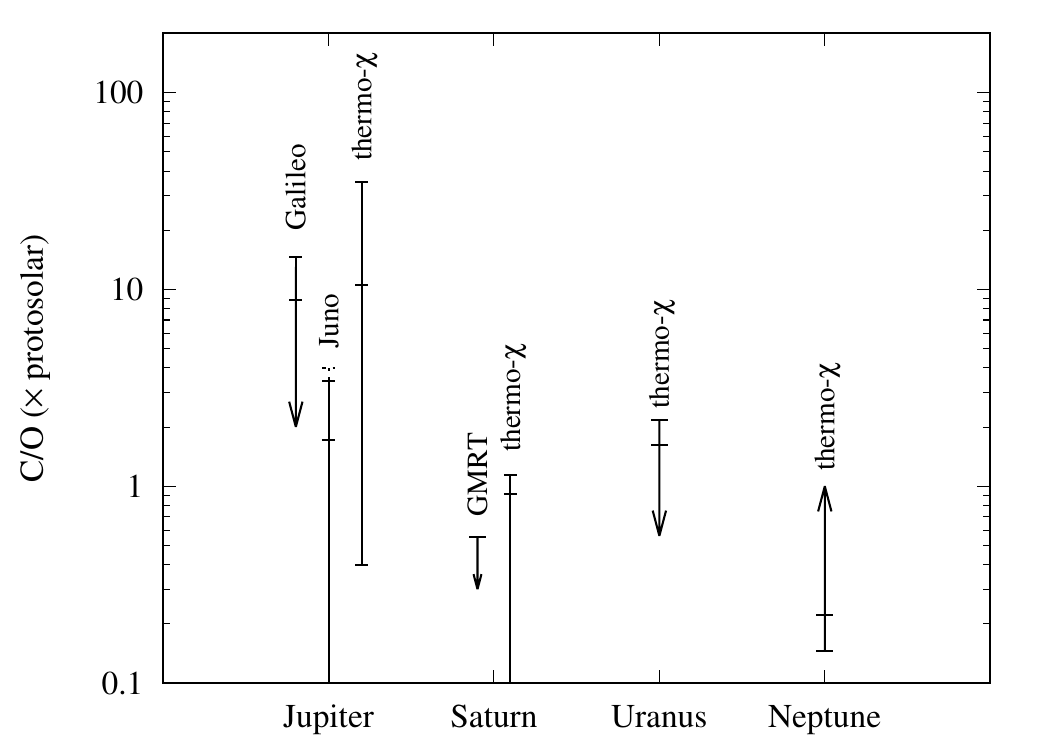}
    \end{center}
    \caption{C/O ratio as derived in the four giant planets using the data from \fig{fig:abundances} (see caption of \fig{fig:abundances} for details) and C abundances taken from Section \ref{Carbon}. The error bars include the uncertainties on the C/H and O/H ratios of the planets and the protosun.}
    \label{fig:CtoO} 
  \end{figure}

  \fig{fig:abundances} summarizes the ranges of O/H ratios in the four giant planets as derived from oxygen species observations and thermochemical models, after rescaling them to the protosolar composition of Lodders (2021). The values displayed on the figure do not account for possible deep radiative layers in Jupiter and Saturn. The deep oxygen abundances would be larger if a radiative layer would be present at the altitude of the CO quenching or deeper (Cavali\'e et al. 2023). For Uranus and Neptune, the radiative layer resulting from the mean molecular weight gradient at the water condensation level is accounted for. This range of values, when combined with the measured abundance of C, results in C/O ratios which are consistent with the solar value, even if the C/O ratio is nominally supersolar for Jupiter and Uranus, and subsolar for Saturn and Neptune, as shown on \fig{fig:CtoO}. However, the O abundance in Uranus is an upper limit and could be subsolar, while it is the reverse situation in Neptune.

  \subsection{Constraints from clouds, storms, and lightning observations}
  Oxygen stored in water condenses at tropospheric levels and water condensation affects the atmospheric dynamics in a number of ways. The atmospheric levels where water can condense depends on the water abundance and the temperature. For water abundances in the range of 0.1 to 3 times solar, water clouds are expected to form at pressure levels which range from less than 3 bar to more than 10 bar in Jupiter, when uncertainties in temperature are included (Wong et al. 2023), and roughly 8-20 bar in Saturn. Water clouds at several tens to hundreds of bars are expected in Uranus and Neptune (Atreya et al. 2020). For Jupiter and Saturn, water clouds should be located about 1.6-3.3 scale heights below the ammonia clouds that cover both planets (i.e. 40-50 km below the visible clouds in Jupiter and 100-130 km below the visible clouds in Saturn), for simple condensation-curve based models. However, direct observations of water clouds in any of these planets is challenging because of the large optical depth of the ammonia clouds. Thus, there are only a handful of observations compatible with water clouds in Jupiter from data acquired either by the Galileo mission (Banfield et al. 1998), or by HST (Wong et al. 2020).  
  The spatial distribution of water in the atmosphere at its condensation level drives changes in the static stability of the atmosphere, influences the behavior of meteorological systems and can manifest more spectacularly in convective storms powered by latent heat (Ingersoll et al. 2004, Sugiyama et al. 2014, Palotai et al. 2022, I\~nurrigarro et al. 2022). We now examine the available information coming from the meteorology of the different Gas and Ice Giants.

  \subsubsection{Jupiter}
  \paragraph{Convective storms}
  Decades of observations of Jupiter show that convective storms make part of the regular activity in the planet (Ingersoll et al. 2004, Vasavada \& Showman 2005). These storms appear with a variety of spatial scales that typically range from one to several thousand kilometers and have active phases with duration from a few hours to weeks in convective events that ocasionally give rise to planetary-scale disturbances (S\'anchez-Lavega et al. 2008, Fletcher et al. 2017). 
  The intensity of the updrafts in the storm can be investigated though the local divergence at the cloud tops (Hunt et al. 1982) and the uppermost levels reached by convective clouds (Stoker 1986, Hueso et al. 2002), especially when combined with numerical models (S\'anchez-Lavega et al. 2008, I\~nurrigarro et al. 2022). The most intense convective storms in Jupiter produce strongly divergent motions with cloud tops near 60$\pm$20 mbar, i.e. above the tropopause. These altitudes are determined from radiative transfer analysis of observations in deeply penetrating atmospheric absorption bands. These overshooting convective storms require energetic moist convection as a power source to overcome the static stability of the upper troposphere. Numerical modeling of the large-scale convective storms in the North Temperate Belt suggest values of deep water abundance of around 3 times solar (S\'anchez-Lavega et al. 2008). Simulating more typical events involving convective storms of lesser intensity requires water abundances ranging from 0.6 to 1.0 times the solar values (I\~nurrigarro et al. 2022). Analysis of the intermittency of convective storms in Jupiter has been used to propose large (10 times solar) abundances of water (Sugiyama et al. 2014).

  \paragraph{Lightning}
  Lightning is abundant in Jupiter's atmosphere. The depth of lightning flashes has been used as a constraint on water abundance suggesting lightning depths at 5 bar and solar water abundance from Voyager observations (Borucki et al. 1986), or much deeper depths from Galileo observations suggesting at least two time solar abundance of water (Dyudina et al. 2002). The study of (Little et al. 1999) suggests lightning depths of 8 bar requiring at least 3 times solar abundances. Numerical models aimed to explain the formation of lightning in Jupiter require at least 0.2 times solar water abundance (Yair et al. 1998). More recently, Aglyamov et al. (2021) used a one-dimensional model of moist convection and lightning generation to investigate the flash rates and the overall characteristics of lightning detected by the Juno mission at a variety of depths, including shallow lightning at around 2 bar (Becker et al. 2020). The model reproduces the variety of depths associated to lightning by considering the antifreeze effect of ammonia dissolved in water. While they find a range of local water abundances, including subsolar values, in which lightning is possible, improved models of deep lightning generation by the same group require at least solar water to fit the data (Aglyamov et al. 2023). 

  \paragraph{Cloud height}
  Spectroscopy can detect cloud top heights, but it is usually not possible to detect the cloud base level if the cloud is thick (Wong et al. 2008). Spectroscopic observations of Jupiter can penetrate in the lower atmosphere in regions with lower cloud content, such as the equatorial dark projections known as hot spots. For instance,  Galileo NIMS spectroscopic images obtained at around 5 $\mu$m were used to infer moist regions of two times solar water abundance in the equatorial region near a dry hot spot (Roos-Serote et al. 2000). Deep clouds have also been constrained by multi-filter imaging (Banfield et al. 1998, Li et al. 2006), but Wong et al. (2023) show that this method has not been able to provide useful constraints on water abundance. Bjoraker et al. (2018) reported a 4-6 bar water cloud inside the Great Red Spot from high resolution infrared spectra obtained in the 5-$\mu$m window. More recently, Bjoraker et al. (2022) used spectroscopic observations to infer the presence of deep clouds between 3 and 7 bar consistent with water clouds requiring 0.5-2.5 times solar abundance of water. Wong et al. (2023) analyzed observations of deep clouds including moist convective storms and used equilibrium condensation cloud models and radiative transfer analysis to examine the spatial distribution of water in the deep clouds. They conclude that a lower limit of O/H $>$ 0.5 protosolar value is needed to explain the observations. \\

  When taken altogether, observations of convective storms, lightning and cloud height indicate that O/H $>$ 0.5 in Jupiter's atmosphere, and tend to favor higher abundances.

  \subsubsection{Saturn}
  Convective storms in Saturn are rarer, possibly of seasonal origin and with most of their activity located in specific latitudinal locations (Dyudina et al. 2007, S\'anchez-Lavega et al. 2020). Planetary-scale disturbances of convective origin occur in Saturn without a clear relation with seasons (S\'anchez-Lavega et al. 2008, 2011, Fletcher et al. 2017). The energetic phenomenology of these storms is inconsistent with ammonia convection (Hueso \& S\'anchez-Lavega 2004), which is much less energetic given the lower latent heat of condensation of water and ammonia (2200-2500 J/g for water depending on the vapor to liquid or vapor to ice phase transition and 1370 J/g for ammonia). Evidence for water ice was found in the clouds of Saturn's Great Storm of 2010-2011, but this did not enable to evaluate the deep O/H (Sromovsky et al. 2013). However, a dynamical analysis of the frequency of giant storms in Saturn by Li \& Ingersoll (2015) suggests values of water abundance of 10 times the solar abundance of Asplund et al. (2009), i.e. 8 times the protosolar abundance of Lodders (2021). Observations of lightning activity well below the ammonia clouds \cite{Dyudina2007} obtained by the Cassini mission are also indicative of deep convection at atmospheric layers where water cloud layers are expected.

  \subsubsection{Uranus and Neptune}
  In the icy giants Uranus and Neptune, water clouds could be located as deep as 50-3000 bar. The high abundance expected for water will have massive effects in the vertical profile of density and temperature of these planets (Hueso et al. 2020). Meteorological features observed in the upper troposphere can be explained by methane and H$_2$S condensation (Lunine 1993, Irwin et al. 2022), and compositional gradients caused by deeper clouds of H$_2$S and NH$_4$SH can be retrieved in microwave observations (Molter et al. 2021, Tollefson et al. 2021, Akins et al. 2023). These observations offer no apparent information about the deep abundance of water coming from the observed meteorology.

\section{Implications on deep atmospheric structure}
\label{Structure}
The deep oxygen abundance, whether it is depleted or enriched, can result in or from rather impactful implications on the internal structure of gas and ice giant planets. We present two different cases in the following section.

  \subsection{Deep radiative layers}
  In the case of Jupiter, a subsolar deep oxygen abundance is consistent with other determinations of O from interior modeling and from MWR data (though at the 2$\sigma$ level), but not from the frequency of deep lightning. Furthermore, a subsolar O abundance implies a C/O value larger than expected from planetesimal accretion, though may be obtained with gas-rich accretion at specific radial distances in the disk (Aguichine et al. 2022).  Quench level models with smaller vertical transport coefficients would predict a higher O, but one cannot arbitrarily reduce K$_{zz}$ if constraints from species like GeH$_4$ are to be respected. One clue to a possibly reduced vigor of vertical mixing in the relevant quench region comes from analysis of the longest wavelength channel of Juno's MWR instrument. Bhattacharya et al. (2023) conclude from this analysis that the abundance of alkali metals in the roughly kilobar pressure region is between 100 and 100,000 times depleted relative to solar. The origin of such a strong depletion remains mysterious. Given such a depletion, the remaining opacity is likely to be not sufficient to maintain a convective environment (Guillot et al. 1994). Instead a radiative layer is expected, in which the vertical mixing could potentially be strongly suppressed, perhaps even to the level of molecular mixing. In this way, the atmosphere is vigorously mixed where K$_{zz}$ indicators such as GeH$_4$ are quenched, but more stagnant where CO quenches, leading to a roughly solar O abundance (Cavali\'e et al. 2023).

  Invocation of a radiative layer raises a number of issues. First is the location and extent of the layer; the pressure level for the top of the layer is close to but not quite consistent with the quench level of CO. To what extent that is adjustable remains an open question. Second, the actual magnitude of vertical mixing in such a layer is highly uncertain. Third, if indeed interior models require a subsolar oxygen, while the oxygen is solar or supersolar in the outer envelope, a radiative layer might be needed to prevent mixing of the two distinct regions. To what extent can quench-level chemistry models be modified to accommodate two very different background elemental abundances above and below the quench zone is also an open question.

  \subsection{Condensation-inhibited convection}
  Guillot (1995) first illustrated that the outer envelopes of giant planets could harbour regions of inhibited convection resulting from strong mean molecular mass gradients. These gradients would be caused by the condensation of species heavier than the hydrogen-helium atmospheres, provided that they would be sufficiently enriched with respect to the solar composition. Leconte et al. (2017) studied the stability of giant planet atmospheres against double-diffusive convection. They found that these regions of transition in mean molecular weight could be radiative with notably steep temperature gradients. They established abundance thresholds beyond which CH$_4$, NH$_3$, H$_2$O and Fe condensation would inhibit convection in giant planets. Cavali\'e et al. (2017) applied their formalism to their thermochemical and diffusion model to constrain the deep oxygen abundance of the ice giants. Their model thus implied that the high oxygen enrichment that was suspected by Lodders \& Fegley (1994), Cavali\'e et al. (2014) and Luszcz-Cook \& de Pater (2013) resulted in a superadiabatic temperature increase in the radiative layer located at the H$_2$O condensation level. This consequently enabled them to decrease the oxygen abundance necessary to reproduce the upper tropospheric CO observations compared to the abovementioned works. These results were confirmed by Venot et al. (2020) with an updated chemical scheme. The deep mass mixing ratio threshold of 9.9 times solar derived by Leconte et al. (2017) for H$_2$O makes it rather unlikely that such process is at play in Jupiter (Li et al. 2020, Cavali\'e et al. 2023), 
  but it may apply to Saturn (see Sections \ref{Saturn} and \ref{Saturn_new}).

\section{Related formation scenarios}
\label{Implications}
To explain Jupiter's supersolar metallicity, including a putative supersolar abundance of oxygen, it has been proposed that its atmosphere could reflect the composition of cold planetesimals made of amorphous ice accreted during the growth of the planet (Owen et al. 1999). Alternative scenarios also predicting a supersolar oxygen abundance in Jupiter and Saturn are those proposing that the two giants accreted solids made of clathrates and/or crystalline ices in their envelopes (Gautier et al. 2001, Hersant et al. 2004, Gautier \& Hersant 2005). Recent thermodynamic and transport models of the PSN based on the presence of pebbles suggest that the combination of disk dynamics with the presence of multiple icelines play a major role into the shaping of the composition of solids formed in the disk (Mousis et al. 2020, Schneeberger et al. 2023). These ice lines are locations where condensation/vaporization cycles can enhance their abundances in both solid and vapor forms. Such models have been used to account for the formation of the N$_2$--rich and H$_2$O--poor Comet C/2016 R2 (PanSTARRS) from solids sharing the same composition and formed in the vicinity of the N$_2$ iceline (Mousis et al. 2021). Similar models have been also utilized to explain the metallicity of giant planets envelopes assuming it was acquired from the accretion of supersolar gases. This includes scenarios where the gas phase in the feeding zone of the giant planets is replenished in volatiles released consecutively to the crystallization of amorphous solids during their inward migration throughout the PSN (Monga \& Desch 2015, Mousis et al. 2019). This also incorporates scenarios proposing that the supersolar gases accreted by Jupiter were delivered by the vaporization of pure condensates during their inward drift through various icelines in the PSN (Aguichine et al. 2022). Finally, the supersolar metallicity of Jupiter's envelope could also result from photoevaporation (Guillot \& Hueso 2006) or from the accretion of enriched gas caused by disk pollution from massive stars in the formation region of the Sun (Throop \& Bally 2008, 2010).

Several formation scenarios have been specifically developed to account for the possibility that Jupiter could be depleted in oxygen. Among them, a carbon--rich Jupiter, which consequently presents a C/O ratio higher than the protosolar value of $\sim$0.5, has been proposed to form between the tar line and the water ice line in the PSN (Lodders 2004).  In this region of the PSN, water is not able to condense but carbonaceous material form, leading the predominance of C--rich solids. Another scenario is that pebbles forming in the vicinity of the CO iceline located in the outer PSN can also present C/O ratios twice higher than the protosolar value. Giant planets formed in this region of the PSN could then be depleted in oxygen relative to carbon (Mousis et al. 2024). Another possibility investigated in the literature is the sequestration of oxygen in refractory material. Even with a protosolar C/O in the gas phase, the abundances of the rock forming elements are sufficient to depress the amount of oxygen available to form water. It has been shown that when the C/O ratio exceeds 0.8, the depletion of oxygen is severe, such that it is bound up in rock, metal oxides, CO and CO$_2$ to the exclusion of water (Pekmezci et al. 2019). Giant planets forming in such an environment would accrete solids devoid in water, and present supersolar C/O ratios.


\section{Conclusion}
\label{Conclusion}
This paper can be summarized as follows:
\begin{itemize}
  \item We have reviewed the various measurements and model results that led to the derivation of the deep oxygen abundance in the giant planets of the solar system. The situation at Jupiter is still unclear, with cloud and spectroscopic observations hinting at supersolar oxygen, while thermochemistry favors subsolar oxygen in the absence of deep radiative layers. Beyond Jupiter, it is not known if the oxygen abundances increases like the carbon abundance, since observations cannot preclude low oxygen values at Uranus and Neptune.
  \item We have presented a new derivation of Saturn's deep oxygen abundance using thermochemical modeling and the CO upper tropospheric abundances derived from VLT/CRIRES observations of Fouchet et al. (2017).
  \item When accounting for the measured carbon abundances, the C/O ratio in the giant planets may be supersolar although this conclusion is tempered by the paucity of direct observational data. For Neptune, the C/O could be subsolar, but this result may only be a lower limit.
\end{itemize}

Deciphering the deep composition of the giant planets is an outstanding task which is necessary to better constrain their internal structure and their formation processes. While the situation is still unclear at Jupiter, the analyses of the Juno/MWR data at the longest wavelength may confirm a supersolar oxygen abundance. At Saturn, and even more so at Uranus and Neptune, combining entry probe and orbiter measurements with new models may help to constrain the deep elemental abundances, even if water may remain inaccessible to a probe. Cavali\'e et al. (2020) proposed to measure the abundances of molecules which can later be related to the deep carriers of the main elements. For instance, the measurements of CO and N$_2$ with a mass spectrometer would help to constrain the deep oxygen and nitrogen when combining these measurements with thermochemical modeling. There are other such examples. Obtaining isotopic measurements would be a valuable complement to help narrow down the main reservoirs of these elements (e.g., N$_2$ or NH$_3$, Fletcher et al. 2014), and thus their formation temperature and heliocentric distance. The lack of data for these planets advocates for sending entry probes (e.g., Mousis et al. 2014, 2018) into their atmospheres, or even multi-probes (Wong et al. 2024)(, with the proviso that reaching the base of the water cloud may be technically extremely challenging. 

\section*{Funding}
T. Cavali\'e and O. Mousis acknowledge support from CNES and the Programme National de Plan\'etologie (PNP) of CNRS/INSU. J. Lunine was supported by the Juno mission through a subcontract from the Southwest Research Institute. R. Hueso was suppported by Grant PID2019-109467GB-I00 funded by MCIN/AEI/10.13039/501100011033/ and by Grupos Gobierno Vasco IT1742-22. The project leading to this publication has received funding from the Excellence Initiative of Aix-Marseille Universit\'e--A*Midex, a French ``Investissements d’Avenir program'' AMX-21-IET-018. This research holds as part of the project FACOM (ANR-22-CE49-0005-01\_ACT) and has benefited from a funding provided by l'Agence Nationale de la Recherche (ANR) under the Generic Call for Proposals 2022. 

\section*{Declarations}
\textbf{Competing Interests} The authors declare no conflict of interest.


\section*{References}
\begin{itemize}
  \item Aglyamov YS, Lunine J, Becker HN, Guillot T, Gibbard SG, Atreya S, Bolton SJ, Levin S, Brown ST, Wong MH (2021) Lightning generation in moist convective clouds and constraints on the water abundance in Jupiter. J Geophys Res, Planets 126(2):e06504. https://doi.org/10.1029/2020JE006504
  \item Aglyamov YS, Lunine J, Atreya S, Guillot T, Becker HN, Levin S, Bolton SJ (2023) Giant planet lightning in nonideal gases. Planet Sci J 4(6):111. https://doi.org/10.3847/PSJ/acd750
  \item Aguichine A, Mousis O, Lunine JI (2022) The possible formation of Jupiter from supersolar gas. Planet Sci J 3(6):141. https://doi.org/10.3847/PSJ/ac6bf1
  \item Akins A, Hofstadter M, Butler B, Friedson AJ, Molter E, Parisi M, de Pater I (2023) Evidence of a polar cyclone on Uranus from VLA observations. Geophys Res Lett 50(10):e2023GL102872. https://doi.org/10.1029/2023GL102872
  \item Asplund M, Grevesse N, Sauval AJ, Scott P (2009) The chemical composition of the Sun. Annu Rev Astron Astrophys 47:481-522. https://doi.org/10.1146/annurev.astro.46.060407.145222
  \item Atreya SK, Wong MH, Owen TC, Mahaffy PR, Niemann HB, de Pater I, Drossart P, Encrenaz T (1999) A comparison of the atmospheres of Jupiter and Saturn: deep atmospheric composition, cloud structure, vertical mixing, and origin. Planet Space Sci 47:1243-1262. https://doi.org/10.1016/S0032-0633(99)00047-1
  \item Atreya SK, Hofstadter MH, In JH, Mousis O, Reh K, Wong MH (2020) Deep atmosphere composition, structure, origin, and exploration, with particular focus on critical in situ science at the icy giants. Space Sci Rev 216(1):18. https://doi.org/10.1007/s11214-020-0640-8
  \item Banfield D, Gierasch PJ, Bell M, Ustinov E, Ingersoll AP, Vasavada AR, West RA, Belton MJS (1998) Jupiter's cloud structure from Galileo imaging data. Icarus 135(1):230-250. https://doi.org/10.1006/icar.1998.5985
  \item Bar-Nun A, Kleinfeld I, Kochavi E (1988) Trapping of gas mixtures by amorphous water ice. Phys Rev B 38:7749-7754. https://doi.org/10.1103/PhysRevB.38.7749
  \item Becker HN, Alexander JW, Atreya SK, Bolton SJ, Brennan MJ, Brown ST, Guillaume A, Guillot T, Ingersoll AP, Levin SM, Lunine JI, Aglyamov YS, Steffes PG (2020) Small lightning flashes from shallow electrical storms on Jupiter. Nature 584(7819):55-58. https://doi.org/10.1038/s41586-020-2532-1
  \item Beer R (1975) Detection of carbon monoxide in Jupiter. Astrophys J Lett 200:L167-L169. https://doi.org/10.1086/181923
  \item B\'ezard B, Lellouch E, Strobel D, Maillard JP, Drossart P (2002) Carbon monoxide on Jupiter: evidence for both internal and external sources. Icarus 159:95-111. https://doi.org/10.1006/icar.2002.6917
  \item Bhattacharya A, Li C, Atreya S et al (2023) Highly depleted alkali metals in Jupiter's deep atmosphere. Astrophys J Lett 952:L27. https://doi.org/10.3847/2041-8213/ace115
  \item Bjoraker GL, Wong MH, de Pater I, Hewagama T, \'ad\'amkovics M, Orton GS (2018) The gas composition and deep cloud structure of Jupiter's great red spot. Astron J 156(3):101. https://doi.org/10.3847/1538-3881/aad186
  \item Bjoraker GL, Wong MH, de Pater I, Hewagama T, \'ad\'amkovics M (2022) The spatial variation of water clouds, NH$_3$, and H$_2$O on Jupiter using keck data at 5 microns. Remote Sens 14(18):4567. https://doi.org/10.3390/rs14184567
  \item Bolton SJ, Adriani A, Adumitroaie V, Allison M, Anderson J, Atreya S, Bloxham J, Brown S, Connerney JEP, DeJong E, Folkner W, Gautier D, Grassi D, Gulkis S, Guillot T, Hansen C, Hubbard WB, Iess L, Ingersoll A, Janssen M, Jorgensen J, Kaspi Y, Levin SM, Li C, Lunine J, Miguel Y, Mura A, Orton G, Owen T, Ravine M, Smith E, Steffes P, Stone E, Stevenson D, Thorne R,Waite J, Durante D, Ebert RW, Greathouse TK, Hue V, Parisi M, Szalay JR, Wilson R (2017) Jupiter's interior and deep atmosphere: the initial pole-to-pole passes with the Juno spacecraft. Science 356(6340):821-825. https://doi.org/10.1126/science.aal2108
  \item BoruckiWJ, Williams MA (1986) Lightning in the Jovian water cloud. J Geophys Res 91:9893-9903. https://doi.org/10.1029/JD091iD09p09893
  \item Cavali\'e T, Billebaud F, Dobrijevic M, Fouchet T, Lellouch E, Encrenaz T, Brillet J, Moriarty-Schieven GH, Wouterloot JGA, Hartogh P (2009) First observation of CO at 345 GHz in the atmosphere of Saturn with the JCMT: new constraints on its origin. Icarus 203:531-540. https://doi.org/10.1016/j.icarus.2009.05.024
  \item Cavali\'e T, Hartogh P, Billebaud F, Dobrijevic M, Fouchet T, Lellouch E, Encrenaz T, Brillet J, Moriarty-Schieven GH (2010) A cometary origin for CO in the stratosphere of Saturn? Astron Astrophys 510:A88
  \item Cavali\'e T, Feuchtgruber H, Lellouch E, de Val-Borro M, Jarchow C, Moreno R, Hartogh P, Orton G, Greathouse TK, Billebaud F, DobrijevicM, Lara LM, Gonz\'alez A, Sagawa H (2013) Spatial distribution of water in the stratosphere of Jupiter from Herschel HIFI and PACS observations. Astron Astrophys 553:A21. https://doi.org/10.1051/0004-6361/201220797
  \item Cavali\'e T, Moreno R, Lellouch E, Hartogh P, Venot O, Orton GS, Jarchow C, Encrenaz T, Selsis F, Hersant F, Fletcher LN (2014) The first submillimeter observation of CO in the stratosphere of Uranus. Astron Astrophys 562:A33. https://doi.org/10.1051/0004-6361/201322297
  \item Cavali\'e T, Venot O, Selsis F, Hersant F, Hartogh P, Leconte J (2017) Thermochemistry and vertical mixing in the tropospheres of Uranus and Neptune. How convection inhibition can affect the derivation of deep oxygen abundances. Icarus 291:1-16
  \item Cavali\'e T, Venot O, Miguel Y, Fletcher LN, Wurz P, Mousis O, Bounaceur R, Hue V, Leconte J, Dobrijevic M (2020) The deep composition of Uranus and Neptune from in situ exploration and thermochemical modeling. Space Sci Rev 216(4):58. https://doi.org/10.1007/s11214-020-00677-8
  \item Cavali\'e T, Lunine J, Mousis O (2023) A subsolar oxygen abundance or a radiative region deep in Jupiter revealed by thermochemical modelling. Nat Astron 7:678-683. https://doi.org/10.1038/s41550-023-01928-8
  \item Connerney JEP, Acuna MH, Ness NF (1987) The magnetic field of Uranus. J Geophys Res 92(A13):15329-15336. https://doi.org/10.1029/JA092iA13p15329
  \item Connerney JEP, Acuna MH, Ness NF (1991) The magnetic field of Neptune. J Geophys Res 96:19023-19042. https://doi.org/10.1029/91JA01165
  \item Connerney JEP, Kotsiaros S, Oliversen RJ, Espley JR, Joergensen JL, Joergensen PS, Merayo JMG, Herceg M, Bloxham J, Moore KM, Bolton SJ, Levin SM (2018) A new model of Jupiter's magnetic field from Juno's first nine orbits. Geophys Res Lett 45(6):2590-2596. https://doi.org/10.1002/2018GL077312
  \item Connerney JEP, Timmins S, Oliversen RJ, Espley JR, Joergensen JL, Kotsiaros S, Joergensen PS, Merayo JMG, Herceg M, Bloxham J, Moore KM, Mura A, Moirano A, Bolton SJ, Levin SM (2022) A new model of Jupiter's magnetic field at the completion of Juno's prime mission. J Geophys Res, Planets 127(2):e07055. https://doi.org/10.1029/2021JE007055
  \item Conrath BJ, Gautier D (2000) Saturn helium abundance: a reanalysis of Voyager measurements. Icarus 144:124-134. https://doi.org/10.1006/icar.1999.6265
  \item Conrath B, Hanel R, Gautier D, Marten A, Lindal G (1987) The helium abundance of Uranus from Voyager measurements. J Geophys Res 92:15003-15010. https://doi.org/10.1029/JA092iA13p15003
  \item Conrath BJ, Gautier D, Lindal GF, Samuelson RE, Shaffer WA (1991) The helium abundance of Neptune from Voyager measurements. J Geophys Res 96:18907
  \item Courtin R, Pandey-Pommier M, Gautier D, Zarka P, Hofstadter M, Hersant F, Girard J (2015) Metric observations of Saturn with the Giant Metrewave Radio Telescope. In: SF2A-2015: Proceedings of the annual meeting of the French society of astronomy and astrophysics, pp 241-245
  \item de Graauw T, Feuchtgruber H, B\'ezard B, Drossart P, Encrenaz T, Beintema DA, Griffin M, Heras A, Kessler M, Leech K, Lellouch E, Morris P, Roelfsema PR, Roos-Serote M, Salama A, Vandenbussche B, Valentijn EA, Davis GR, Naylor DA (1997) First results of ISO-SWS observations of Saturn: detection of CO$_2$, CH$_3$C$_2$H, C$_4$H$_2$ and tropospheric H$_2$O. Astron Astrophys 321:L13-L16
  \item de Pater I, Dickel JR (1991) Multifrequency radio observations of Saturn at ring inclination angles between 5 and 26 degrees. Icarus 94(2):474-492. https://doi.org/10.1016/0019-1035(91)90242-L
de Pater I, Richmond M (1989) Neptune's microwave spectrum from 1 mm to 20 cm. Icarus 80:1-13. https://doi.org/10.1016/0019-1035(89)90158-9
  \item de Pater I, Romani PN, Atreya SK (1991) Possible microwave absorption by H2S gas in Uranus' and Neptune's
atmospheres. Icarus 91:220-233. https://doi.org/10.1016/0019-1035(91)90020-T
  \item de Pater I, Molter EM, Moeckel CM (2023) A review of radio observations of the giant planets: probing the composition, structure, and dynamics of their deep atmospheres. Remote Sens 15(5):1313. https://doi.org/10.3390/rs15051313
  \item DoughertyMK, Cao H, Khurana KK, Hunt GJ, Provan G, Kellock S, Burton ME, Burk TA, Bunce EJ, Cowley SWH, Kivelson MG, Russell CT, Southwood DJ (2018) Saturn's magnetic field revealed by the Cassini grand finale. Science 362(6410):aat5434. https://doi.org/10.1126/science.aat5434
  \item Dunn DE, de Pater I, Wright M, Hogerheijde MR, Molnar LA (2005) High-quality BIMA-OVRO images of Saturn and its rings at 1.3 and 3 millimeters. Astron J 129:1109-1116. https://doi.org/10.1086/424536
  \item Dyudina UA, Ingersoll AP, Vasavada AR, Ewald SP, Galileo SSI Team (2002) Monte Carlo radiative transfer modeling of lightning observed in Galileo images of Jupiter. Icarus 160(2):336-349. https://doi.org/10.1006/icar.2002.6977
  \item Dyudina UA, Ingersoll AP, Ewald SP, Porco CC, Fischer G, Kurth W, Desch M, Del Genio A, Barbara J, Ferrier J (2007) Lightning storms on Saturn observed by Cassini ISS and RPWS during 2004 2006. Icarus 190(2):545-555. https://doi.org/10.1016/j.icarus.2007.03.035
  \item Encrenaz T, de Graauw T, Schaeidt S, Lellouch E, Feuchtgruber H, Beintema DA, B\'ezard B, Drossart P, Griffin M, Heras A, Kessler M, Leech K, Morris P, Roelfsema PR, Roos-Serote M, Salama A, Vandenbussche B, Valentijn EA, Davis GR, Naylor DA (1996) First results of ISO-SWS observations of Jupiter. Astron Astrophys 315:L397-L400
  \item Encrenaz T, Lellouch E, Drossart P, Feuchtgruber H, Orton GS, Atreya SK (2004) First detection of CO in Uranus. Astron Astrophys 413:L5-L9. https://doi.org/10.1051/0004-6361:20034637
  \item Fegley B, Prinn RG (1988) Chemical constraints on the water and total oxygen abundances in the deep atmosphere of Jupiter. Astrophys J 324:621-625. https://doi.org/10.1086/165922
  \item Feuchtgruber H, Lellouch E, Orton G, de Graauw T, Vandenbussche B, Swinyard B, Moreno R, Jarchow C, Billebaud F, Cavali\'e T, Sidher S, Hartogh P (2013) The D/H ratio in the atmospheres of Uranus and Neptune from Herschel-PACS observations. Astron Astrophys 551:A126. https://doi.org/10.1051/0004-6361/201220857
  \item Fletcher LN, Irwin PGJ, Teanby NA, Orton GS, Parrish PD, Calcutt SB, Bowles N, de Kok R, Howett C, Taylor FW (2007) The meridional phosphine distribution in Saturn's upper troposphere from Cassini/CIRS observations. Icarus 188:72-88. https://doi.org/10.1016/j.icarus.2006.10.029
  \item Fletcher LN, Orton GS, Teanby NA, Irwin PGJ, Bjoraker GL (2009) Methane and its isotopologues on Saturn from Cassini/CIRS observations. Icarus 199:351-367. https://doi.org/10.1016/j.icarus.2008.09.019
  \item Fletcher LN, Greathouse TK, Orton GS, Irwin PGJ, Mousis O, Sinclair JA, Giles RS (2014) The origin of nitrogen on Jupiter and Saturn from the 15N/14N ratio. Icarus 238:170
  \item Fletcher LN, Orton GS, Rogers JH, Giles RS, Payne AV, Irwin PGJ, Vedovato M (2017) Moist convection and the 2010-2011 revival of Jupiter's south equatorial belt. Icarus 286:94
  \item Folkner WM, Iess L, Anderson JD, Asmar SW, Buccino DR, Durante D, Feldman M, Gomez Casajus L, Gregnanin M, Milani A, Parisi M, Park RS, Serra D, Tommei G, Tortora P, Zannoni M, Bolton SJ, Connerney JEP, Levin SM (2017) Jupiter gravity field estimated from the first two Juno orbits. Geophys Res Lett 44(10):4694-4700. https://doi.org/10.1002/2017GL073140
  \item Fouchet T, Lellouch E, Cavali\'e T, B\'ezard B (2017) First determination of the tropospheric CO abundance in Saturn. In: AAS/division for planetary sciences meeting abstracts, vol 49, p 209.05
  \item Friedson AJ, Gonzales EJ (2017) Inhibition of ordinary and diffusive convection in the water condensation zone of the ice giants and implications for their thermal evolution. Icarus 297:160-178. https://doi.org/10.1016/j.icarus.2017.06.029
  \item Gautier D, Hersant F (2005) Formation and composition of planetesimals. Space Sci Rev 116:25-52. https://doi.org/10.1007/s11214-005-1946-2
  \item Gautier D, Conrath B, Flasar M, Hanel R, Kunde V, Chedin A, Scott N (1981) The helium abundance of Jupiter from Voyager. J Geophys Res 86(A10):8713-8720. https://doi.org/10.1029/JA086iA10p08713
  \item Gautier D, Hersant F, Mousis O, Lunine JI (2001) Enrichments in volatiles in Jupiter: a new interpretation of the Galileo measurements. Astrophys J Lett 550:L227-L230. https://doi.org/10.1086/319648
  \item Guillot T (1995) Condensation of methane, ammonia, and water and the inhibition of convection in giant planets. Science 269:1697-1699. https://doi.org/10.1126/science.7569896
  \item Guillot T (2005) The interiors of giant planets: models and outstanding questions. Annu Rev Earth Planet Sci 33:493-530. https://doi.org/10.1146/annurev.earth.32.101802.120325. arXiv:astro-ph/0502068
  \item Guillot T, Hueso R (2006) The composition of Jupiter: sign of a (relatively) late formation in a chemically evolved protosolar disc. Mon Not R Astron Soc 367(1):L47-L51. https://doi.org/10.1111/j.1745-3933.2006.00137.x. arXiv:astro-ph/0601043
  \item Guillot T, Gautier D, Chabrier G, Mosser B (1994) Are the giant planets fully convective? Icarus 112(2):337-353. https://doi.org/10.1006/icar.1994.1188
  \item Guillot T, Fletcher LN, Helled R, Ikoma M, Line MR, Parmentier V (2022) Giant Planets from the Inside-Out. arXiv e-prints arXiv:2205.04100. https://doi.org/10.48550/arXiv.2205.04100
  \item Hanel R, Conrath B, Herath L, Kunde V, Pirraglia J (1981) Albedo, internal heat, and energy balance of Jupiter - preliminary results of the Voyager infrared investigation. J Geophys Res 86:8705-8712. https://doi.org/10.1029/JA086iA10p08705
  \item Hanel RA, Conrath BJ, Kunde VG, Pearl JC, Pirraglia JA (1983) Albedo, internal heat flux, and energy balance of Saturn. Icarus 53:262-285. https://doi.org/10.1016/0019-1035(83)90147-1
  \item Helled R, Fortney JJ (2020) The interiors of Uranus and Neptune: current understanding and open questions. Philos Trans R Soc Lond Ser A 378(2187):20190474. https://doi.org/10.1098/rsta.2019.0474
  \item Helled R, Lunine J (2014) Measuring Jupiter's water abundance by Juno: the link between interior and formation models. Mon Not R Astron Soc 441:2273-2279. https://doi.org/10.1093/mnras/stu516
  \item Helled R, Anderson JD, Podolak M, Schubert G (2011) Interior models of Uranus and Neptune. Astrophys J 726:15. https://doi.org/10.1088/0004-637X/726/1/15
  \item Helled R, Nettelmann N, Guillot T (2020) Uranus and Neptune: origin, evolution and internal structure. Space Sci Rev 216(3):38. https://doi.org/10.1007/s11214-020-00660-3
  \item Hersant F, Gautier D, Lunine JI (2004) Enrichment in volatiles in the giant planets of the solar system. Planet Space Sci 52:623-641. https://doi.org/10.1016/j.pss.2003.12.011
  \item Hesman BE, Davis GR, Matthews HE, Orton GS (2007) The abundance profile of CO in Neptune's atmosphere. Icarus 186(2):342-353. https://doi.org/10.1016/j.icarus.2006.08.025
  \item Hueso R, S\'anchez-Lavega A (2004) A three-dimensional model of moist convection for the giant planets II: Saturn's water and ammonia moist convective storms. Icarus 172(1):255-271. https://doi.org/10.1016/j.icarus.2004.06.010
  \item Hueso R, S\'anchez-Lavega A, Guillot T (2002) A model for large-scale convective storms in Jupiter. J Geophys Res, Planets 107(E10):5075. https://doi.org/10.1029/2001JE001839
  \item Hueso R, Guillot T, S\'anchez-Lavega A (2020) Convective storms and atmospheric vertical structure in Uranus and Neptune. Philos Trans R Soc Lond Ser A 378(2187):20190476. https://doi.org/10.1098/rsta.2019.0476
  \item Hunt GE, Muller JP, Gee P (1982) Convective growth rates of equatorial features in the Jovian atmosphere. Nature 295(5849):491-494. https://doi.org/10.1038/295491a0
  \item Iess L, Militzer B, Kaspi Y, Nicholson P, Durante D, Racioppa P, Anabtawi A, Galanti E, HubbardW, Mariani MJ, Tortora P, Wahl S, Zannoni M (2019) Measurement and implications of Saturn's gravity field and ring mass. Science 364(6445):aat2965. https://doi.org/10.1126/science.aat2965
  \item Ingersoll AP, Dowling TE, Gierasch PJ, Orton GS, Read PL, S\'anchez-Lavega A, Showman AP, Simon-Miller AA, Vasavada AR (2004) Dynamics of Jupiter's atmosphere. In: Bagenal F, Dowling TE, McKinnon WB (eds) Jupiter: the planet, satellites and magnetosphere. Cambridge University Press, pp 105-128
  \item I\~nurrigarro P, Hueso R, S\'anchez-Lavega A, Legarreta J (2022) Convective storms in closed cyclones in Jupiter: (II) numerical modeling. Icarus 386:115169. https://doi.org/10.1016/j.icarus.2022.115169
  \item Irwin PGJ, Toledo D, Garland R, Teanby NA, Fletcher LN, Orton GA, B\'ezard B (2018) Detection of hydrogen sulfide above the clouds in Uranus's atmosphere. Nat Astron 2:420-427. https://doi.org/10.1038/s41550-018-0432-1
  \item Irwin PGJ, Toledo D, Braude AS, Bacon R, Weilbacher PM, Teanby NA, Fletcher LN, Orton GS (2019a) Latitudinal variation in the abundance of methane (CH$_4$) above the clouds in Neptune's atmosphere from VLT/MUSE narrow field mode observations. Icarus 331:69
  \item Irwin PGJ, Toledo D, Garland R, Teanby NA, Fletcher LN, Orton GS, B\'ezard B (2019b) Probable detection of hydrogen sulphide (H$_2$S) in Neptune's atmosphere. Icarus 321:550-563. https://doi.org/10.1016/j.icarus.2018.12.014
  \item Irwin PGJ, Dobinson J, James A, Toledo D, Teanby NA, Fletcher LN, Orton GS, P\'erez-Hoyos S (2021) Latitudinal variation of methane mole fraction above clouds in Neptune's atmosphere from VLT/MUSENFM: limb-darkening reanalysis. Icarus 357:114277. https://doi.org/10.1016/j.icarus.2020.114277
  \item Irwin PGJ, Teanby NA, Fletcher LN, Toledo D, Orton GS, Wong MH, Roman MT, P\'erez-Hoyos S, James A, Dobinson J (2022) Hazy blue worlds: a holistic aerosol model for Uranus and Neptune, including dark spots. J Geophys Res, Planets 127(6):e07189. https://doi.org/10.1029/2022JE007189
  \item Karkoschka E, Tomasko M (2009) The haze and methane distributions on Uranus from HST-STIS spectroscopy. Icarus 202:287-309. https://doi.org/10.1016/j.icarus.2009.02.010
  \item Karkoschka E, Tomasko MG (2011) The haze and methane distributions on Neptune from HST-STIS spectroscopy. Icarus 211:780-797. https://doi.org/10.1016/j.icarus.2010.08.013
  \item Koskinen TT, Guerlet S (2018) Atmospheric structure and helium abundance on Saturn from Cassini/UVIS and CIRS observations. Icarus 307:161-171. https://doi.org/10.1016/j.icarus.2018.02.020
  \item Kunde V, Hanel R, Maguire W, Gautier D, Baluteau JP, Marten A, Chedin A, Husson N, Scott N (1982) The tropospheric gas composition of Jupiter's North equatorial belt /NH$_3$, PH$_3$, CH$_3$D, GeH$_4$, H$_2$O/ and the Jovian D/H isotopic ratio. Astrophys J 263:443-467. https://doi.org/10.1086/160516
  \item Larson HP, Fink U, Treffers R, Gautier TN III (1975) Detection of water vapor on Jupiter. Astrophys J 197:L137-L140
  \item Leconte J, Selsis F, Hersant F, Guillot T (2017) Condensation-inhibited convection in hydrogen-rich atmospheres. Stability against double-diffusive processes and thermal profiles for Jupiter, Saturn, Uranus, and Neptune. Astron Astrophys 598:A98. https://doi.org/10.1051/0004-6361/201629140
  \item Lellouch E, Paubert G, Moreno R, Festou MC, B\'ezard B, Bockelee-Morvan D, Colom P, Crovisier J, Encrenaz T, Gautier D,Marten A, Despois D, Strobel DF, Sievers A (1995) Chemical and thermal response of Jupiter's atmosphere following the impact of comet Shoemaker-Levy-9. Nature 373:592-595. https://doi.org/10.1038/373592a0
  \item Lellouch E, B\'ezard B, Fouchet T, Feuchtgruber H, Encrenaz T, de Graauw T (2001) The deuterium abundance in Jupiter and Saturn from ISO-SWS observations. Astron Astrophys 370:610-622. https://doi.org/10.1051/0004-6361:20010259
  \item Lellouch E, B\'ezard B, Moses JI, Davis GR, Drossart P, Feuchtgruber H, Bergin EA, Moreno R, Encrenaz T (2002) The origin of water vapor and carbon dioxide in Jupiter's stratosphere. Icarus 159:112-131. https://doi.org/10.1006/icar.2002.6929
  \item Lellouch E, Moreno R, Paubert G (2005) A dual origin for Neptune's carbon monoxide? Astron Astrophys 430:L37-L40. https://doi.org/10.1051/0004-6361:200400127
  \item Li C, Ingersoll AP (2015) Moist convection in hydrogen atmospheres and the frequency of Saturn's giant storms. Nat Geosci 8(5):398-403. https://doi.org/10.1038/ngeo2405
  \item Li L, Ingersoll AP, Vasavada AR, Simon-Miller AA, Del Genio AD, Ewald SP, Porco CC, West RA (2006) Vertical wind shear on Jupiter from Cassini images. J Geophys Res, Planets 111(E4):E04004. https://doi.org/10.1029/2005JE002556
  \item Li C, Ingersoll A, Janssen M, Levin S, Bolton S, Adumitroaie V, Allison M, Arballo J, Bellotti A, Brown S, Ewald S, Jewell L, Misra S, Orton G, Oyafuso F, Steffes P, Williamson R (2017) The distribution of ammonia on Jupiter from a preliminary inversion of Juno microwave radiometer data. Geophys Res Lett 44(11):5317-5325. https://doi.org/10.1002/2017GL073159
  \item Li C, Ingersoll A, Bolton S, Levin S, Janssen M, Atreya S, Lunine J, Steffes P, Brown S, Guillot T, Allison M, Arballo J, Bellotti A, Adumitroaie V, Gulkis S, Hodges A, Li L, Misra S, Orton G, Oyafuso F,
Santos-Costa D,Waite H, Zhang Z (2020) The water abundance in Jupiter's equatorial zone. Nat Astron 4:609-616. https://doi.org/10.1038/s41550-020-1009-3
  \item Li C, Allison MD, Atreya SK, Fletcher LN, Galanti E, Guillot T, Ingersoll AP, Kaspi Y, Li L, Lunine JI, Orton G, Oyafuso FA, Steffes PG,Wong MH, Zhang Z, Levin S, Bolton SJ (2022) Jupiter's tropospheric temperature and composition. In: AGU fall meeting abstracts, vol 2022, pp P32C-1854
  \item Li C, de Pater I, Moeckel C, Sault RJ, Butler B, deBoer D, Zhang Z (2023a) Long-lasting, deep effect of Saturn's giant storms. Sci Adv 9(32):eadg9419. https://doi.org/10.1126/sciadv.adg9419
  \item Li C, Allison M, Atreya S, Fletcher L, Ingersoll A, Li L, Orton G, Oyafuso F, Steffes P, Wong M, Zhang Z, Levin S, Bolton S (2023b) A new value of Jupiter's deep isentrope - implications for Jupiter's deep thermal and compositional structure. In: EGU general assembly conference abstracts, EGU general assembly conference abstracts, pp EGU-10747. https://doi.org/10.5194/egusphere-egu23-10747
  \item Little B, Anger CD, Ingersoll AP, Vasavada AR, Senske DA, Breneman HH, Borucki WJ, Galileo SSI Team (1999) Galileo images of lightning on Jupiter. Icarus 142(2):306-323. https://doi.org/10.1006/icar.1999.6195
  \item Lodders K (2004) Jupiter formed with more tar than ice. Astrophys J 611:587-597. https://doi.org/10.1086/421970
  \item Lodders K (2021) Relative atomic solar system abundances, mass fractions, and atomic masses of the elements and their isotopes, composition of the solar photosphere, and compositions of the major chondritic meteorite groups. Space Sci Rev 217(3):44. https://doi.org/10.1007/s11214-021-00825-8
  \item Lodders K, Fegley B Jr (1994) The origin of carbon monoxide in Neptunes's atmosphere. Icarus 112:368-375. https://doi.org/10.1006/icar.1994.1190
  \item Lunine JI (1993) The atmospheres of Uranus and Neptune. Annu Rev Astron Astrophys 31:217-263. https://doi.org/10.1146/annurev.aa.31.090193.001245
  \item Lunine JI, Stevenson DJ (1985) Thermodynamics of clathrate hydrate at low and high pressures with application to the outer solar system. Astrophys J Suppl 58:493-531. https://doi.org/10.1086/191050
  \item Luszcz-Cook SH, de Pater I (2013) Constraining the origins of Neptune's carbon monoxide abundance with CARMA millimeter-wave observations. Icarus 222(1):379-400. https://doi.org/10.1016/j.icarus.2012.11.002
  \item Mahaffy PR, Niemann HB, Alpert A, Atreya SK, Demick J, Donahue TM, Harpold DN, Owen TC (2000) Noble gas abundance and isotope ratios in the atmosphere of Jupiter from the Galileo probe mass spectrometer. J Geophys Res 105:15061-15072. https://doi.org/10.1029/1999JE001224
  \item Marten A, Gautier D, Owen T, Sanders DB, Matthews HE, Atreya SK, Tilanus RPJ, Deane JR (1993) First observations of CO and HCN on Neptune and Uranus at millimeter wavelengths and the implications for atmospheric chemistry. Astrophys J 406:285-297. https://doi.org/10.1086/172440
  \item Molter EM, de Pater I, Luszcz-Cook S, Tollefson J, Sault RJ, Butler B, de Boer D (2021) Tropospheric composition and circulation of Uranus with ALMA and the VLA. Planet Sci J 2(1):3. https://doi.org/10.3847/PSJ/abc48a
  \item Monga N, Desch S (2015) External photoevaporation of the solar nebula: Jupiter's noble gas enrichments. Astrophys J 798(1):9. https://doi.org/10.1088/0004-637X/798/1/9
  \item Moses JI (2014) Chemical kinetics on extrasolar planets. Philos Trans R Soc Lond Ser A 372:20130073-20130073. https://doi.org/10.1098/rsta.2013.0073
  \item Moses JI, Poppe AR (2017) Dust ablation on the giant planets: consequences for stratospheric photochemistry. Icarus 297:33
  \item Mousis O, Lunine JI, Madhusudhan N, Johnson TV (2012) Nebular water depletion as the cause of Jupiter's low oxygen abundance. Astrophys J Lett 751:L7. https://doi.org/10.1088/2041-8205/751/1/L7
  \item Mousis O, Fletcher LN, Lebreton JP, Wurz P, Cavali\'e T, Coustenis A, Courtin R, Gautier D, Helled R, Irwin PGJ, Morse AD, Nettelmann N, Marty B, Rousselot P, Venot O, Atkinson DH, Waite JH, Reh KR, Simon AA, Atreya S, Andr\'e N, Blanc M, Daglis IA, Fischer G, Geppert WD, Guillot T, Hedman MM, Hueso R, Lellouch E, Lunine JI, Murray CD, O'Donoghue J, Rengel M, S\'anchez-Lavega A, Schmider FX, Spiga A, Spilker T, Petit JM, Tiscareno MS, Ali-Dib M, Altwegg K, Bolton SJ, Bouquet A, Briois C, Fouchet T, Guerlet S, Kostiuk T, Lebleu D, Moreno R, Orton GS, Poncy J (2014) Scientific rationale for Saturn's in situ exploration. Planet Space Sci 104:29-47. https://doi.org/10.1016/j.pss.2014.09.014
  \item Mousis O, Atkinson DH, Cavali\'e T, Fletcher LN, Amato MJ, Aslam S, Ferri F, Renard JB, Spilker T, Venkatapathy E, Wurz P, Aplin K, Coustenis A, Deleuil M, Dobrijevic M, Fouchet T, Guillot T, Hartogh P, Hewagama T, Hofstadter MD, Hue V, Hueso R, Lebreton JP, Lellouch E, Moses J, Orton GS, Pearl JC, S\'anchez-Lavega A, Simon A, Venot O, Waite JH, Achterberg RK, Atreya S, Billebaud F, Blanc M, Borget F, Brugger B, Charnoz S, Chiavassa T, Cottini V, d'Hendecourt L, Danger G, Encrenaz T, Gorius NJP, Jorda L,Marty B, Moreno R, Morse A, Nixon C, Reh K, Ronnet T, Schmider FX, Sheridan S, Sotin C, Vernazza P, Villanueva GL (2018) Scientific rationale for Uranus and Neptune in situ explorations. Planet Space Sci 155:12-40. https://doi.org/10.1016/j.pss.2017.10.005
  \item Mousis O, Ronnet T, Lunine JI (2019) Jupiter's formation in the vicinity of the amorphous ice snowline. Astrophys J 875(1):9. https://doi.org/10.3847/1538-4357/ab0a72
  \item Mousis O, Aguichine A, Helled R, Irwin PGJ, Lunine JI (2020) The role of ice lines in the formation of Uranus and Neptune. Philos Trans R Soc Lond Ser A 378(2187):20200107. https://doi.org/10.1098/rsta.2020.0107
  \item Mousis O, Aguichine A, Bouquet A, Lunine JI, Danger G, Mandt KE, Luspay-Kuti A (2021a) Cold traps of hypervolatiles in the protosolar nebula at the origin of the peculiar composition of comet C/2016 R2 (PanSTARRS). Planet Sci J 2(2):72. https://doi.org/10.3847/PSJ/abeaa7
  \item Mousis O, Lunine JI, Aguichine A (2021b) The nature and composition of Jupiter's building blocks derived from the water abundance measurements by the Juno spacecraft. Astrophys J Lett 918(2):L23. https://doi.org/10.3847/2041-8213/ac1d50
  \item Mousis O, Cavali\'e T, Lunine JI, Mandt KE, Hueso R, Aguichine A, Schneeberger A, Benest Couzinou T, Atkinson DH, Hue V, Hofstadter M, Srisuchinwong U (2024) Recipes for forming a carbon-rich giant planet. Space Sci Rev 220
  \item Movshovitz N, Fortney JJ (2022) The promise and limitations of precision gravity: application to the interior structure of Uranus and Neptune. Planet Sci J 3(4):88. https://doi.org/10.3847/PSJ/ac60ff
  \item Ness NF, Acuna MH, Burlaga LF, Connerney JEP, Lepping RP, Neubauer FM (1989) Magnetic fields at Neptune. Science 246(4936):1473-1478. https://doi.org/10.1126/science.246.4936.1473
  \item Nettelmann N, Helled R, Fortney JJ, Redmer R (2013) New indication for a dichotomy in the interior structure of Uranus and Neptune from the application of modified shape and rotation data. Planet Space Sci 77:143-151. https://doi.org/10.1016/j.pss.2012.06.019
  \item Nettelmann N, Wang K, Fortney JJ, Hamel S, Yellamilli S, Bethkenhagen M, Redmer R (2016) Uranus evolution models with simple thermal boundary layers. Icarus 275:107
  \item Niemann HB, Atreya SK, Carignan GR, Donahue TM, Haberman JA, Harpold DN, Hartle RE, Hunten DM, Kasprzak WT, Mahaffy PR, Owen TC, Way SH (1998) The composition of the Jovian atmosphere as determined by the Galileo probe mass spectrometer. J Geophys Res 103:22831-22846. https://doi.org/10.1029/98JE01050
  \item Nixon CA, Irwin PGJ, Calcutt SB, Taylor FW, Carlson RW (2001) Atmospheric composition and cloud structure in Jovian 5- ?m hotspots from analysis of Galileo NIMS measurements. Icarus 150(1):48-68. https://doi.org/10.1006/icar.2000.6561
  \item Noll KS, Larson HP (1991) The spectrum of Saturn from 1990 to 2230 cm ?1: abundances of AsH$_3$, CH$_3$D, CO, GeH$_4$, NH$_3$, and PH$_3$. Icarus 89(1):168-189. https://doi.org/10.1016/0019-1035(91)90096-C
  \item Noll KS, Knacke RF, Geballe TR, Tokunaga AT (1986) Detection of carbon monoxide in Saturn. Astrophys J Lett 309:L91-L94. https://doi.org/10.1086/184768
  \item Orton GS, Fisher BM, Baines KH, Stewart ST, Friedson AJ, Ortiz JL, Marinova M, Ressler M, Dayal A, Hoffmann W, Hora J, Hinkley S, Krishnan V, Masanovic M, Tesic J, Tziolas A, Parija KC (1998)
Characteristics of the Galileo probe entry site from Earth-based remote sensing observations. J Geophys Res 103(E10):22791-22814. https://doi.org/10.1029/98JE02380
  \item Owen T, Mahaffy P, Niemann HB, Atreya S, Donahue T, Bar-Nun A, de Pater I (1999) A low-temperature origin for the planetesimals that formed Jupiter. Nature 402:269-270. https://doi.org/10.1038/46232
  \item Pacetti E, Turrini D, Schisano E, Molinari S, Fonte S, Politi R, Hennebelle P, Klessen R, Testi L, Lebreuilly U (2022) Chemical diversity in protoplanetary disks and its impact on the formation history of giant planets. Astrophys J 937(1):36. https://doi.org/10.3847/1538-4357/ac8b11
  \item Palotai C, Brueshaber S, Sankar R, Sayanagi K (2022) Moist convection in the giant planet atmospheres. Remote Sens 15(1):219. https://doi.org/10.3390/rs15010219
  \item Pearl JC, Conrath BJ (1991) The albedo, effective temperature, and energy balance of Neptune, as determined from Voyager data. J Geophys Res 96:18921
  \item Pearl JC, Conrath BJ, Hanel RA, Pirraglia JA (1990) The albedo, effective temperature, and energy balance of Uranus, as determined from Voyager IRIS data. Icarus 84:12-28. https://doi.org/10.1016/0019-1035(90)90155-3
  \item Pekmezci GS, Johnson TV, Lunine JI, Mousis O (2019) A statistical approach to planetesimal condensate composition beyond the snowline based on the carbon-to-oxygen ratio. Astrophys J 887(1):3. https://doi.org/10.3847/1538-4357/ab4c4a
  \item Prinn RG, Barshay SS (1977) Carbon monoxide on Jupiter and implications for atmospheric convection. Science 198:1031-1034. https://doi.org/10.1126/science.198.4321.1031
  \item Roos-Serote M, Vasavada AR, Kamp L, Drossart P, Irwin P, Nixon C, Carlson RW (2000) Proximate humid and dry regions in Jupiter's atmosphere indicate complex local meteorology. Nature
405(6783):158-160. https://doi.org/10.1038/35012023
  \item Rosenqvist J, Lellouch E, Romani PN, Paubert G, Encrenaz T (1992) Millimeter-wave observations of Saturn, Uranus, and Neptune - CO and HCN on Neptune. Astrophys J Lett 392:L99-L102. https://doi.org/10.1086/186435
  \item S\'anchez-Lavega A, Orton GS, Hueso R, Garc\'ia-Melendo E, P\'erez-Hoyos S, Simon-Miller A, Rojas JF, G\'omez JM, Yanamandra-Fisher P, Fletcher L, Joels J, Kemerer J, Hora J, Karkoschka E, de Pater I, Wong MH, Marcus PS, Pinilla-Alonso N, Carvalho F, Go C, Parker D, Salway M, Valimberti M, Wesley A, Pujic Z (2008) Depth of a strong Jovian jet from a planetary-scale disturbance driven by storms. Nature 451(7177):437-440. https://doi.org/10.1038/nature06533
  \item S\'anchez-Lavega A, del R\'io-Gaztelurrutia T, Hueso R, G\'omez-Forrellad JM, Sanz-Requena JF, Legarreta J, Garc\'ia-Melendo E, Colas F, Lecacheux J, Fletcher LN, Barrado y Navascu\'es D, Parker D, International Outer Planet Watch Team, Akutsu T, Barry T, Beltran J, Buda S, Combs B, Carvalho F, Casquinha P, Delcroix M, Ghomizadeh S, Go C, Hotershall J, Ikemura T, Jolly G, Kazemoto A, Kumamori T, Lecompte M, Maxson P, Melillo FJ, Milika DP, Morales E, Peach D, Phillips J, Poupeau JJ, Sussenbach J, Walker G, Walker S, Tranter T, Wesley A, Wilson T, Yunoki K (2011) Deep winds beneath Saturn's upper clouds from a seasonal long-lived planetary-scale storm. Nature 475:71-74. https://doi.org/10.1038/nature10203
  \item S\'anchez-Lavega A, Garc\'ia-Melendo E, Legarreta J, Hueso R, del R\'io-Gaztelurrutia T, Sanz-Requena JF, P\'erez-Hoyos S, Simon AA, Wong MH, Soria M, G\'omez-Forrellad JM, Barry T, Delcroix M, Sayanagi KM, Blalock JJ, Gunnarson JL, Dyudina U, Ewald S (2020) A complex storm system in Saturn's North polar atmosphere in 2018. Nat Astron 4:180-187. https://doi.org/10.1038/s41550-019-0914-9
  \item Schneeberger A, Mousis O, Aguichine A, Lunine JI (2023) Evolution of the reservoirs of volatiles in the protosolar nebula. Astron Astrophys 670:A28. https://doi.org/10.1051/0004-6361/202244670
  \item Schneider AD, Bitsch B (2021) How drifting and evaporating pebbles shape giant planets. II. Volatiles and refractories in atmospheres. Astron Astrophys 654:A72. https://doi.org/10.1051/0004-6361/202141096
  \item Showman AP, Dowling TE (2000) Nonlinear simulations of Jupiter's 5-micron hot spots. Science 289(5485):1737-1740. https://doi.org/10.1126/science.289.5485.1737
  \item Sromovsky LA, Fry PM (2008) The methane abundance and structure of Uranus' cloud bands inferred from spatially resolved 2006 Keck grism spectra. Icarus 193:252-266. https://doi.org/10.1016/j.icarus.2007.08.037
  \item Sromovsky LA, Collard AD, Fry PM, Orton GS, Lemmon MT, Tomasko MG, Freedman RS (1998) Galileo probe measurements of thermal and solar radiation fluxes in the Jovian atmosphere. J Geophys Res 103(E10):22929-22978. https://doi.org/10.1029/98JE01048
  \item Sromovsky LA, Fry PM, Kim JH (2011) Methane on Uranus: the case for a compact CH$_4$ cloud layer at low latitudes and a severe CH 4 depletion at high-latitudes based on re-analysis of Voyager occultation measurements and STIS spectroscopy. Icarus 215:292
  \item Sromovsky LA, Baines KH, Fry PM (2013) Saturn's great storm of 2010-2011: evidence for ammonia and water ices from analysis of VIMS spectra. Icarus 226(1):402-418. https://doi.org/10.1016/j.icarus.2013.05.043
  \item Sromovsky LA, Karkoschka E, Fry PM, Hammel HB, de Pater I, Rages K (2014) Methane depletion in both polar regions of Uranus inferred from HST/STIS and Keck/NIRC2 observations. Icarus 238:137-155. https://doi.org/10.1016/j.icarus.2014.05.016
  \item Stoker CR (1986) Moist convection: a mechanism for producing the vertical structure of the Jovian equatorial plumes. Icarus 67(1):106-125. https://doi.org/10.1016/0019-1035(86)90179-X
  \item Sugiyama K, Nakajima K, Odaka M, Kuramoto K, Hayashi YY (2014) Numerical simulations of Jupiter's moist convection layer: structure and dynamics in statistically steady states. Icarus 229:71-91. https://doi.org/10.1016/j.icarus.2013.10.016
  \item Teanby NA, Irwin PGJ (2013) An external origin for carbon monoxide on Uranus from Herschel/spire? Astrophys J Lett 775(2):L49
  \item Teanby NA, Irwin PGJ, Moses JI (2019) Neptune's carbon monoxide profile and phosphine upper limits from Herschel/SPIRE: implications for interior structure and formation. Icarus 319:86-98. https://doi.org/10.1016/j.icarus.2018.09.014
  \item Teanby NA, Irwin PGJ, Moses JI, Helled R (2020) Neptune and Uranus: ice or rock giants? Philos Trans R Soc Lond Ser A 378(2187):20190489. https://doi.org/10.1098/rsta.2019.0489
  \item Throop HB, Bally J (2008) Tail-end bondi-hoyle accretion in young star clusters: implications for disks, planets, and stars. Astron J 135(6):2380-2397. https://doi.org/10.1088/0004-6256/135/6/2380
  \item Throop HB, Bally J (2010) Accretion of Jupiter's atmosphere from a supernova-contaminated molecular cloud. Icarus 208(1):329-336. https://doi.org/10.1016/j.icarus.2010.02.005
  \item Tollefson J, de Pater I, Molter EM, Sault RJ, Butler BJ, Luszcz-Cook S, DeBoer D (2021) Neptune's spatial brightness temperature variations from the VLA and ALMA. Planet Sci J 2(3):105. https://doi.org/10.3847/PSJ/abf837
  \item Vasavada AR, Showman AP (2005) Jovian atmospheric dynamics: an update after Galileo and Cassini. Rep Prog Phys 68(8):1935-1996. https://doi.org/10.1088/0034-4885/68/8/R06
  \item Venot O, H\'ebrard E, Ag\'undez M, Dobrijevic M, Selsis F, Hersant F, Iro N, Bounaceur R (2012) A chemical model for the atmosphere of hot Jupiters. Astron Astrophys 546:A43. https://doi.org/10.1051/0004-6361/201219310
  \item Venot O, Cavali\'e T, Bounaceur R, Tremblin P, Brouillard L, Lhoussaine Ben Brahim R (2020) New chemical scheme for giant planet thermochemistry. Update of the methanol chemistry and new reduced chemical scheme. Astron Astrophys 634:A78. https://doi.org/10.1051/0004-6361/201936697
  \item Visscher C, Fegley B Jr (2005) Chemical constraints on the water and total oxygen abundances in the deep atmosphere of Saturn. Astrophys J 623:1221-1227. https://doi.org/10.1086/428493. arXiv:astroph/0501128
  \item Visscher C, Moses JI (2011) Quenching of carbon monoxide and methane in the atmospheres of cool brown dwarfs and hot jupiters. Astrophys J 738:72. https://doi.org/10.1088/0004-637X/738/1/72
  \item Visscher C, Moses JI, Saslow SA (2010) The deep water abundance on Jupiter: new constraints from thermochemical kinetics and diffusion modeling. Icarus 209:602
  \item von Zahn U, Hunten DM (1992) The Jupiter helium interferometer experiment on the Galileo entry probe. Space Sci Rev 60(1-4):263-281. https://doi.org/10.1007/BF00216857
  \item von Zahn U, Hunten DM, Lehmacher G (1998) Helium in Jupiter's atmosphere: results from the Galileo probe helium interferometer experiment. J Geophys Res 103:22815-22830. https://doi.org/10.1029/98JE00695
  \item Wang D, Gierasch PJ, Lunine JI, Mousis O (2015) New insights on Jupiter's deep water abundance from disequilibrium species. Icarus 250:154
  \item Wang D, Lunine JI,Mousis O (2016) Modeling the disequilibrium species for Jupiter and Saturn: implications for Juno and Saturn entry probe. Icarus 276:21
  \item Wong MH, Mahaffy PR, Atreya SK, Niemann HB, Owen TC (2004) Updated Galileo probe mass spectrometer measurements of carbon, oxygen, nitrogen, and sulfur on Jupiter. Icarus 171:153-170. https://doi.org/10.1016/j.icarus.2004.04.010
  \item Wong MH, Lunine JI, Atreya SK, Johnson T, Mahaffy PR, Owen TC, Encrenaz T (2008) Oxygen and other volatiles in the giant planets and their satellites. Rev Mineral Geochem 68(1):219-246. https://doi.org/10.2138/rmg.2008.68.10
  \item Wong MH, Simon AA, Tollefson JW, de Pater I, Barnett MN, Hsu AI, Stephens AW, Orton GS, Fleming SW, Goullaud C, Januszewski W, Roman A, Bjoraker GL, Atreya SK, Adriani A, Fletcher LN (2020) High-resolution UV/Optical/IR Imaging of Jupiter in 2016-2019. Astrophys J Suppl 247(2):58. https://doi.org/10.3847/1538-4365/ab775f
  \item Wong MH, Bjoraker GL, Goullaud C, Stephens AW, Luszcz-Cook SH (2023) Deep clouds on Jupiter. Remote Sens 15(3):702. https://doi.org/10.3390/rs15030702
  \item Wong MH, Markham S, Rowe-Gurney N, Sayanagi K, Hueso R (2024) Multiple probe measurements at Uranus motivated by spatial variability. Space Sci Rev 220
  \item Yair Y, Levin Z, Tzivion S (1998) Model interpretation of Jovian lightning activity and the Galileo probe results. J Geophys Res 103(D12):14157-14166. https://doi.org/10.1029/98JD00310
  \item Yung YL, Drew WA, Pinto JP, Friedl RR (1988) Estimation of the reaction rate for the formation of CH$_3$O from H $+$ H$_2$CO: implications for chemistry in the solar system. Icarus 73:516-526. https://doi.org/10.1016/0019-1035(88)90061-9      
  
\end{itemize}

\end{document}